\documentstyle[12pt]{article}
\begin{document}






\def\dsp{\displaystyle}
\def\Rr{{bf R}}
\def\Zz{bf Z}
\def\Nn{bf N}
\def\get{\hbox{{\goth g}$^*$}}
\def\g{\gamma}
\def\om{\omega}
\def\r{\rho}
\def\a{\alpha}
\def\s{\sigma}
\def\vfi{\varphi}
\def\l{\lambda}
\def\implique{\Rightarrow}
\def\o{{\circ}}
\def\Diff{\hbox{\rm Diff}}
\def\S1{\hbox{\rm S$^1$}}
\def\Hom{\hbox{\rm Hom}}
\def\Vect{\hbox{\rm Vect}}
\def\const{\hbox{\rm const}}
\def\ad{\hbox{\hbox{\rm ad}}}
\def\semid{\hbox{\bb o}}
\def\blanc{\hbox{\ \ }}

\def\pds#1,#2{\langle #1\mid #2\rangle} 
\def\f#1,#2,#3{#1\colon#2\to#3} 

\def\hfl#1{{\buildrel{#1}\over{\hbox to
12mm{\rightarrowfill}}}}

\title{Space of second order linear differential operators as a module over
the Lie algebra of vector fields}

\author{C. Duval, V.Yu. Ovsienko
\thanks{C.N.R.S., Centre de Physique Th\'eorique, Luminy-Case 907, F-13288
Marseille Cedex 9, France}}

\maketitle

\abstract{The space of linear differential operators on a smooth manifold $M$
has a natural one-parameter family of $\Diff (M)$ (and $\Vect (M)$)-module
structures, defined by their action on the space of tensor densities. It is
shown that, in the case of second order differential operators, the $\Vect
(M)$-module structures are equivalent for any degree of tensor-densities
except for three critical values: $\{0,{1\over 2},1\}$. Second order analogue
of the Lie derivative appears as an intertwining operator between the spaces
of second order differential operators on tensor-densities.}

\vskip 3cm
\hskip 7cm
hep-th/9409065

\bigskip
{\bf Keywords:} Differential operators, tensor-densities, cohomology
of vector fields

\vfill\eject

\section{Introduction: main problem}

Let $M$ be an oriented manifold of dimension $n$. Consider the space ${\cal
D}^k(M)$ of $k$-th order linear differential operators on $M$. In local
coordinates, such an operator is given by:
\begin{equation}
A(\phi)
=
a_k^{{i_1}\ldots{i_k}}\partial_{i_1}\ldots\partial_{i_k}\phi
+
\cdots
+
a_1^i\partial_i\phi
+
a_0\phi
\label{1}
\end{equation}
where $\partial_i=\frac{\partial}{\partial x^i}$ and
$a_k^{{i_1}\ldots{i_l}},\phi\in C^{\infty}(M)$
with $l=0,1,\ldots,k$.
(From now on we suppose a summation over repeated indices.)

The group $\Diff (M)$ of all diffeomorphisms of $M$ and the Lie algebra
$\Vect (M)$ of all smooth vector fields naturally act on the space
${\cal D}^k(M)$. Let $G\in \Diff (M)$, then the action is defined by
$$
G(A):=G^{*-1}AG^*.
\label{2}
$$
A vector field $\xi \in \Vect (M)$ acts on differential operators by the
commutator with the operator of Lie derivative:
\begin{equation}
\ad L_{\xi}(A):=L_{\xi}\circ A-A\circ L_{\xi}
\label{3}
\end{equation}

It is interesting to take as arguments {\it tensor-densities} of
degree $\lambda$ instead of functions. This defines a family of $\Diff (M)$ and
$\Vect (M)$-module structures on ${\cal D}^k(M)$ depending on $\lambda$.

\vskip 0,3cm

Studying these module structures is a very important problem since a number of
different examples naturally appear in differential geometry and
mathematical physics (see below). To our knowledge, the classification problem
of such module structures for different values of $\lambda$ has never been
considered (at least in the multidimensional case). In this paper we solve
this problem for the space ${\cal D}^2(M)$ of second order linear
differential operators.

\subsection {Tensor-densities: definition}

Consider the determinant bundle $\Lambda ^nTM\rightarrow M$. The group
${\bf R}^*$ acts on the fibers by multiplication.

\proclaim Definition.
A homogeneous function of degree $\lambda$ on the complement
$\Lambda^n TM\setminus M$ of the zero section of the determinant bundle:
$$
F(\kappa w)=\kappa ^{\lambda}F(w)
$$
is called tensor-density of degree $\lambda$ on $M$.\par

\goodbreak

Let us denote ${\cal F}_{\lambda}(M)$ the space of tensor-densities of degree
$-\lambda$. It is evident that ${\cal F}_0(M)=C^{\infty}(M)$, the space ${\cal
F}_{-1}(M)$ coincides with the space of differential $n$-forms on $M$:
${\cal F}_{-1}(M)=\Omega^n(M)$.

In local coordinates, one uses the following notation for a tensor-density of
degree $\lambda$:
$$
\phi =\phi (x_1,...,x_n)(dx_1\wedge ... \wedge dx_n)^{\lambda }.
 \label{4}
$$
The group $\Diff (M)$ acts on the determinant bundle by homogeneous
diffeomorphisms. Therefore, it acts on the space ${\cal F}_{\lambda}$. One
has:
$$
G^*\phi =\phi \circ G^{-1}\cdot J_G^{\lambda}
$$
where $J_G=\frac{DG}{Dx}$ is the Jacobian.

\goodbreak

The corresponding action of the Lie algebra $\Vect (M)$ is given by the Lie
derivative:
\begin{equation}
L_{\xi}\phi =\xi^i\partial_i\phi -\lambda \phi \partial_i\xi^i
\label{5}
\end{equation}
Remark that this formula does not depend on the choice of coordinates.

It is evident that, for an oriented manifold,
${\cal F}_{\lambda}\cong {\cal F}_{\mu}$ as linear
spaces (but not as modules) for any $\lambda,\mu$.

{\bf Remark}. If $M$ is compact, then there exists a natural isomorphism of
$\Vect (M)$- (and $\Diff (M)$-) modules
$$
{\cal F}_{\lambda}\cong {\cal F}_{-1-\lambda}
$$
(tautological for $\lambda =-{1\over 2}$). Indeed, there exists a
non-degenerate
invariant pairing ${\cal F}_{\lambda}\otimes {\cal F}_{-1-\lambda}\rightarrow
{\bf R}$ given by $\langle\phi,\psi\rangle =\int_M\phi\psi$.

\subsection{Linear differential operators on 
tensor-densities}

\proclaim Definition. Each vector
field $\xi$ defines an operator $L_{\xi}$ on the space of tensor-densities
${\cal F}_{\lambda}$. Define the space ${\cal D}^k_{\lambda}$ of $k$-th order
linear differential operators on ${\cal F}_{\lambda}$ as the space of all
$k$-th order polynomials in different operators $L_{\xi}$ and operators of
multiplication by functions.\par

Each differential operator $A\in {\cal D}^k_{\lambda}$ is given by (\ref{1})
in any system of local coordinates.

{\bf Examples}. 1) A classical example is the theory of Sturm-Liouville
equation:  $\phi''(x)+u(x)\phi (x)=0$. The argument $\phi $, in this case, is a
$-{1\over2}$-density: $\phi = \phi (x)(dx)^{-{1\over 2}}$
(see \cite{car,kir1}).

2) Another example is given by geometric quantization
\cite{sou,kos1}: the algebra of linear differential operators acting on
a Hilbert space of ${1\over 2}$-densities \cite{bla,kos2}
(see also \cite{sni,kir2,woo}).

\vskip 0,3cm

The space of differential operators on ${\cal F}_{\lambda}$ does not depend
on $\lambda$ as a linear space. We shall use the notation ${\cal
D}^k_{\lambda}$ for $\Diff (M)$ and $\Vect (M)$-module structures on this
space.

\subsection{The Lie derivative as a $\Vect (M)$-equi\-variant operator}

All the modules ${\cal D}^1_{\lambda}$ of first order differential operators
are isomorphic to each other: there exists a {\it $\Vect (M)$-equivariant}
linear mapping
$$
{\cal L}_{\mu\;\lambda}:{\cal D}^1_{\mu}\rightarrow {\cal
D}^1_{\lambda}.
$$
Let $A=a_1^i\partial_i+a_0\;\in {\cal D}^1_{\mu}$,
define
\begin{equation}
{\cal L}_{\mu\;\lambda}(A)=a_1^i\partial_i+a_0+(\mu -\lambda )\partial
_ia_1^i
\label{6}
\end{equation}

In fact,
the expression $A(\phi)$ can be written in invariant way:
$A(\phi )=L_{a_1}\phi +(a_0+\mu \partial_ia_1^i)\phi $
(the quantity $a_0+\mu \partial_ia_1^i$ transforms as a function by coordinate
transformations).
The operator ${\cal L}_{\mu\;\lambda}(A)$ is defined by
a simple application of the same operator to tensor-densities of degree
$\lambda$. Let $\psi
\in {\cal F}_{\lambda }$, put ${\cal L}_{\mu\;\lambda}(A)\psi:=L_{a_1}\psi
+(a_0+\mu \partial_ia_1^i)\psi $. One obtains the explicit formula (\ref{6})
from (\ref{5}).

 The mapping ${\cal L}_{\mu\;\lambda}$ generalizes the classical Lie
derivative. In fact,
$$
L_{\xi}={\cal L}_{0\;\lambda}(\xi).
$$
More generally, if $L_{\xi}^{(\lambda )}$ is the Lie derivative on
${\cal F}_{\lambda }$, then ${\cal L}_{\mu\;\lambda}(L_{\xi}^{(\mu )})=
L_{\xi}^{(\lambda )}$.

We are now looking for a higher-order analogue of the
Lie derivative, in other words, for an equivariant linear
mapping
$$
{\cal L}^k_{\mu\;\lambda}:{\cal D}^k_{\mu}\rightarrow {\cal D}^k_{\lambda}.
$$

{\bf Remark}. Let us recall that a classic theorem (see \cite{rud,kir3}) states
that the only $\Vect (M)$-equivariant differential operator in {\it one}
argument ({\it unar} operator) on the space of `geometric quantities'
(tensors, tensor-densities, etc.) is the standard differential of functions:
$d:C^{\infty}(M)\rightarrow\Omega^1(M)$. Now, the space of differential
operators turns out to be a richer module, so that the Lie derivative (\ref{6})
is admitted as a unar equivariant differential operator. The purpose of this
paper is to find another equivariant operator extending the ordinary Lie
derivative.

\section{Main theorems}

The main results of this paper corresponds to the space ${\cal D}^2$ of second
order linear differential operators on an oriented manifold $M$:
$$
A(\phi )=a_2^{ij}\partial_i\partial_j\phi +a_1^i\partial_i\phi +a_0\phi.
$$
The formula (\ref{3}) defines a family of $\Vect (M)$-module structures
${\cal D}^2_{\lambda }$ on this space.

\subsection{Critical values of the degree}

 The following theorem gives the
classification of the $\Vect (M)$- modules ${\cal D}^2_{\lambda }$ on the space
of second order linear differential operators.

\proclaim Theorem 1. (i) If $\dim M\geq 2$, then all $\Vect (M)$- modules
${\cal D}^2_{\lambda }$
with $\lambda \not =0,-{1\over 2},-1$ are isomorphic
to each other, but not isomorphic to
$$
{\cal D}^2_0\;\cong \;{\cal D}^2_{-1}\;\not\cong\;{\cal
D}^2_{-{1\over 2}}.
$$
(ii) If $\dim M=1$, then all $\Vect (M)$-modules
${\cal D}^2_{\lambda }$
with $\lambda \not =0,-1$ are isomorphic to each other, but not isomorphic
to
$$
{\cal D}^2_0 \;\cong \;{\cal D}^2_{-1}.
$$

Therefore, there is one stable $\Vect (M)$-module structure on
the space of second order linear differential operators
and two exceptional modules corresponding to functions and to $1\over
2$-densities, if $\dim M\geq 2$. If $\dim M=1$, there are only two different
$\Vect (M)$-module structures.

\vskip 0,3cm

\proclaim Definition. Let us call critical values the following values of the
degree : $\{0,{1\over 2},1\}$ if  $\dim M\geq 2$ and $\{0,1\}$ if
$\dim M=1$.\par

\vskip 0,3cm

\subsection{Second order Lie derivative}

We propose
here an explicit formula for the equivariant linear mapping
$$
{\cal L}^2_{\mu\;\lambda}:{\cal D}^2_{\mu}\rightarrow {\cal D}^2_{\lambda}
$$
which can be considered as an analogue of the Lie derivative.
Let us introduce the following notation:
$\widetilde A={\cal L}^2_{\mu\;\lambda}(A)$.

\proclaim Theorem 2. (i) $\dim M \geq 2$.
Let $\lambda ,\;\mu \not =0,-{1\over 2},-1$.
There exists a unique (up to a constant) equivariant linear mapping
$
{\cal L}^2_{\mu\;\lambda}:{\cal D}^2_{\mu}\rightarrow {\cal D}^2_{\lambda}
$
given by the formula:
\begin{equation}
\left\{
\begin{array}{rcl}
\widetilde a_2^{ij}
&=&
a_2^{ij} \\ \noalign{\smallskip}
\widetilde a_1^l
&=&
{\dsp \frac{2\lambda +1}{2\mu+1}}\,a_1^l
+2{\dsp \frac{\mu-\lambda }{2\mu +1}}\,\partial_ia_2^{il} \\
\noalign{\smallskip}
\widetilde a_0
&=&
{\dsp \frac{\lambda (\lambda +1)}{\mu (\mu +1)}}\,a_0
+{\dsp \frac{\lambda(\mu -\lambda)}{(2\mu +1)(\mu +1)}}\,
(\partial_ia_1^{i}-\partial_i\partial_ja_2^{ij})
\end{array}
\right.
\label{8}
\end{equation}
(ii) $\dim M =1$. Let $\lambda,\mu\not =0,-1$. There exists
a 1-parameter family of equivariant linear mappings from ${\cal D}^2_{\mu}$
to ${\cal D}^2_{\lambda}$.\par

As a result of equivariance, the formula (\ref{8}) does not depend on the
choice of
coordinates.

{\bf Remark}. If $\lambda =\mu$, then $\widetilde A=A$; if $\lambda
+\mu=-1$, then the mapping (\ref{8}) is the conjugation of differential
operators: $\widetilde A=A^*$. Thus, the mapping (\ref{8}) realizes an
interpolation between a differential operator and its conjugate.

\subsection{Hierarchy of modules}

For the critical values of the degree $-\lambda $, the $\Vect (M)$-module
${\cal D}^2_{\lambda}$ is not isomorphic to ${\cal D}^2_{\mu}$
(with general $\mu $). The mapping
${\cal L}^2_{\mu\;\lambda}$ in this case, is an {\it invariant projections}
from
${\cal D}^2_{\mu}$ to ${\cal D}^2_{\lambda}$. The following diagram represents
the hierarchy of $\Vect (M)$-module structures on the space of second order
differential operators:
$$
\matrix{
&&{\cal D}^2_{\mu}
&&&&
{\cal D}^2_{\mu}\cr
&\swarrow
&&
\searrow
&&&\downarrow \cr
\hfill{\cal D}^2_{-{1\over 2}}
&&&&{\cal D}^2_0 \cong {\cal D}^2_{-1} \hfill
&&{\cal D}^2_{0}\cong{\cal D}^2_{-1} \cr
&&&&&&\cr
&&
\dim M\geq 2
&& &&
\dim M=1 \cr
&&&&&\kern 1,5cm& \cr
}
$$

\goodbreak

\subsection{Automorphisms of exceptional modules}

Recall that a linear mapping
${\cal I}:{\cal D}^2_{\lambda}\rightarrow {\cal D}^2_{\lambda}$ is called an
{\it automorphism} of the $\Vect (M)$-module ${\cal D}^2_{\lambda}$ if
$$
[{\cal I},\ad L_{\xi}]=0
$$
for any $\xi\in\Vect (M)$.

An important property of the $\Vect (M)$-module structures on ${\cal D}^2$
corresponding to critical values of the degree, is the existence of
non-trivial automorphisms.

The uniqueness of the mapping (\ref{8}) implies the following fact:

\proclaim Corollary of Theorem 2. The $\Vect (M)$-module ${\cal
D}^2_{\lambda}$ with $\lambda \not =0,\allowbreak -{1\over 2},-1$ has no
nontrivial automorphisms.\par

The following statement gives the classification of automorphisms for the
modules  ${\cal D}^2_0\cong {\cal D}^2_{-1}$ and ${\cal D}^2_{-{1\over 2}}$.

\proclaim Proposition 1.
(i) Each automorphism of the module ${\cal D}^2_0$ is
proportional to the following one:
\begin{equation}
{\cal I}
\left(
\matrix{
a_2 \cr
a_1 \cr
a_0 \cr
} \right)= \left(
\matrix{
 a_2 \hfill \cr
 a_1 \hfill \cr
c\cdot a_0 \cr
} \right)
\label{auto0}
\end{equation}
where $c=\const$.
\hfill\break
(ii) Each automorphism of the module ${\cal D}^2_{-{1\over 2}}$ is
proportional to the following one:
\begin{equation}
{\cal I} \left(
\matrix{
a_2^{ij} \cr\noalign{\smallskip}
a_1^i \cr\noalign{\smallskip}
a_0 \cr\noalign{\smallskip}
} \right)= \left(
\matrix{
a_2^{ij} \hfill \cr\noalign{\smallskip}
a_1^i -2c(a_1^i-\partial_ja_2^{ij})\hfill \cr\noalign{\smallskip}
a_0 -c(\partial_ia_1^i-\partial_i\partial_ja_2^{ij})\cr\noalign{\smallskip}
}
\right)
\label{auto}
\end{equation}
where $c=\const$,
in terms of the components $a_2,a_1,a_0$ of the operator
$A=a_2^{ij}\partial_i\partial_j +a_1^i\partial_i +a_0$.

\goodbreak

\vskip 0,5cm

The next part of the paper is devoted to proofs. We give the explicit
formul{\ae} for the $\Vect (M)$-actions on the space of second order linear
differential operators. This family of actions, depending on $\lambda $,
can be considered as a {\it one-parameter deformation} of a $\Vect (M)$-module
structure.  This approach leads to the cohomology of the Lie algebra $\Vect
(M)$ with some nontrivial operator coefficients.

\section{Action of the Lie algebra $\Vect (M)$ on the space of operators}

The spaces of second order linear differential operators ${\cal D}^2$ is
isomorphic, as a vector space, to a direct sum of some spaces of tensor fields:
$$
{\cal D}^2_{\lambda }\cong S^2(M)\oplus \Vect (M)\oplus C^{\infty }(M)
$$
where $S^2(M)$ is the space of second order symmetric contravariant tensor
fields.

The $\Vect (M)$-action (\ref{3}) on
the space of differential operators
${\cal D}^2_{\lambda }$
is therefore, a `modification' of the standard $\Vect (M)$-action on this
direct sum.

\subsection{Explicit formul{\ae}}

\proclaim Lemma 1. The action $\ad L_{\xi}=\ad L^{(\lambda )}_{\xi }$
of $\Vect (M)$ on ${\cal D}^2_{\lambda }$ (see (\ref{3}))
is given by
\begin{equation}
\left\{
\matrix{
\ad L_{\xi }(A)_2^{ij}
&=&
(L_{\xi }a_2)^{ij} \hfill\cr\noalign{\smallskip}
\ad L_{\xi }(A)_1^l\hfill
&=&
(L_{\xi }a_1)^l-a_2^{ij}\partial_i\partial_j\xi^l
+2\lambda a_2^{li}\partial_i\partial_j\xi^j\hfill\cr\noalign{\smallskip}
\ad L_{\xi }(A)_0\hfill
&=&
L_{\xi }a_0+\lambda
(a_1^r\partial_r
+a_2^{ij}\partial_i\partial_j)\partial_k\xi^k\hfill\cr\noalign{\smallskip}
}\right.
\label{9}
\end{equation}
where
$
(L_{\xi }a_2)^{ij}
=
\xi^r\partial_ra_2^{ij}-a_2^{rj}\partial_r\xi^i-a_2^{ri}\partial_r\xi^j
$ and
$
(L_{\xi }a_1)^l
=
\xi ^r\partial_ra_1^l-a_1^r\partial_r\xi^l
$
and
$
L_{\xi }a_0
=
\xi^r\partial_r a_0
$
are the Lie derivatives of tensor
fields along the vector field $\xi $.\par

{\bf Proof}. By definition, the result of the action
$\ad L_{\xi }$ is given by
$\ad L_{\xi }(A)(\phi )=[L_{\xi },A](\phi )$. From (\ref{3}) one has:
$$
\matrix{
[L_{\xi },A](\phi)
&=&
\xi^r\partial_r(a_2^{ij}\partial_i\partial_j\phi +
a_1^i\partial_i\phi +a_0\phi)\hfill\cr\noalign{\smallskip}
&&-\lambda \partial_k(\xi^k)(a_2^{ij}\partial_i\partial_j\phi
+ a_1^i\partial_i\phi+a_0\phi)\hfill\cr\noalign{\smallskip}
&&-(a_2^{ij}\partial_i\partial_j+a_1^i\partial_i+a_0)(\xi ^r\partial_r\phi
-\lambda \partial_k(\xi ^k)\phi).\hfill\cr\noalign{\smallskip}
}
$$
One gets immediately the formula (\ref{9}).

\vskip 0,3cm

{\bf Remark}.
The homogeneous part of the operator $A$ transforms as a symmetric
contravariant 2-tensor : $a_2\in S^2 (M)$.

\vskip 0,3cm

\subsection{Invariant normal form}

There exists a canonical form of the $\Vect (M)$-action on the space ${\cal
D}^2_{\lambda}$. To obtain it, we introduce here some `transformation of
variables' in the space ${\cal D}^2_{\lambda}$.

The intuitive idea is as follows. First, remark that
the transformation law of the quantity $a_0+\lambda\partial_ia_1^i$ contains
only terms with $a_2$. Second, the action (\ref{9}) is a deformation of the
standard action of $\Vect (M)$ on the space
$S^2(M)\oplus \Vect (M)\oplus
C^{\infty }(M)$. We are therefore trying to find some canonical form of the
cocycles on $\Vect (M)$ generating this deformation.

\vskip 0,3cm
{\bf A}. $\dim M\geq 2$.
\vskip 0,3cm

\proclaim Lemma 2. The following quantities:
\begin{equation}
\left\{
\matrix{
\bar a_2^{ij} &=& a_2^{ij} \hfill\cr\noalign{\smallskip}
\bar a_1^l \hfill &=& a_1^l  +2\lambda
\partial_ia_2^{li}\hfill\cr\noalign{\smallskip}
\bar a_0 \hfill &=&a_0 +\lambda \partial_ia_1^i+{\lambda }^2\partial
_i\partial_ja_2^{ij}\hfill\cr\noalign{\smallskip}
}\right.
\label{10}
\end{equation}
are transformed by the $\Vect (M)$-action in the following way:
\begin{equation}
\left\{
\matrix{
\ad L_{\xi }({\bar a}_2)^{ij}
&=&
\xi ^k\partial_k{\bar a}_2^{ij}-
{\bar a}_2^{ik}\partial_k\xi ^j-{\bar
a}_2^{kj}\partial_k\xi^i\hfill\cr\noalign{\smallskip}
\ad L_{\xi }({\bar a}_1)^i\hfill
&=&
\xi ^k\partial_k{\bar a}_1^i-{\bar a}_1^k\partial_k\xi^i
-(2\lambda +1){\bar
a}_2^{kj}\partial_k\partial_j\xi^i\hfill\cr\noalign{\smallskip}
\ad L_{\xi }({\bar a}_0)\hfill
&=&
\xi ^k\partial_k{\bar a}_0-\lambda
(\lambda +1)\partial_i({\bar
a}_2^{jk})\partial_j\partial_k\xi^i\hfill\cr\noalign{\smallskip}
}\right.
\label{11}
\end{equation}\par

{\bf Proof}:
 From the formula (\ref{9}), one has: $\partial_kL_{\xi}a_2^{ki}=
\xi ^r\partial_r\partial_ka_2^{ki}-\partial_ka_2^{kr}\partial_r\xi^i-
a_2^{kr}\partial_k\partial_r\xi^i-a_2^{ir}\partial_r\partial_k\xi ^k$.
The transformation law for ${\bar a}_1$ in the formula (\ref{10}) follows
immediately from this expression.

The transformation law for ${\bar a}_0$ can be easily verified in the same
way.

\proclaim Important remark. The mapping
$$
\gamma :
\xi ^r\partial_r\mapsto \partial_i\partial_j \xi ^r\cdot dx^i\otimes
dx^j\otimes \partial_r
\label{coc}
$$
is a nontrivial 1-cocycle on $\Vect (M)$ with values in the space of
2-covariant, 1-contravariant tensors. This cocycle appears in the Lie
derivative of connections (see Sec.~6).\par

\goodbreak

{\bf Remark}.
The formula (\ref{11}) can be written in more invariant way. It is sufficient
to
note that the terms depending on ${\bar a}_2$ in the expressions
$\ad L_{\xi}({\bar a}_1)$ and $\ad L_{\xi }({\bar a}_0)$ are respectively equal
to
$-(2\lambda +1)\langle {\bar a}_2,\gamma (\xi)\rangle$ and
$-\lambda (\lambda +1)\big[\partial_i\langle{\bar a}_2,\gamma (\xi)\rangle^i -
{\bar a}_2(\partial_i\xi^i)\big]$, where
${\bar a}_2(f):={\bar a}_2^{ij}\partial_j\partial_kf$.

\vskip 0,3cm

{\bf B}. $\dim M=1$.

\vskip 0,3cm

In the one-dimensional case, let us simply give the normal form.

\proclaim Lemma 3. The following quantities:
$$
\left\{
\begin{array}{rcl}
\bar a_2 &=& a_2 \\ \noalign{\smallskip}
\bar a_1 &=& a_1  +{1\over 2}(2\lambda -1)a_2' \\ \noalign{\smallskip}
\bar a_0 &=& a_0 +\lambda a_1'+{1\over 3}\lambda (2\lambda -1)a_2''
\end{array}
\right.
$$
transform under the action of $\Vect (M)$ according to:
$$
\left\{
\matrix{
\ad L_{\xi }({\bar a}_2)
&=&
\xi {\bar a}_2'-2\xi '{\bar a}_2 \hfill \cr\noalign{\smallskip}
\ad L_{\xi }({\bar a}_1)
&=&
\xi {\bar a}_1'-\xi '{\bar a}_1 \hfill  \cr\noalign{\smallskip}
\ad L_{\xi }({\bar a}_0)
&=&
\xi {\bar a}_0'+{2\over 3}\lambda(\lambda +1){\bar a}_2\xi''' \hfill
}\right.
$$\par

\section{Proof of Theorem 1}

First of all, it is evident that all the modules ${\cal D}^2_{\lambda}$ with
$\lambda \not =0,-{1\over 2},-1$ are isomorphic. In fact, the mapping
$A\mapsto \widetilde A$ given in normal coordinates (\ref{10}) by:
$\widetilde {\bar a}_2=\bar a_2,\;
\widetilde {\bar a}_1=\frac{2\mu +1}{2\lambda +1}\bar a_1,\;
\widetilde {\bar a}_0=\frac{\mu (\mu +1)}{\lambda (\lambda +1)}{\bar a}_0$
defines an isomorphism between ${\cal D}^2_{\lambda}$ and
${\cal D}^2_{\mu}$.

In the same way, one has: ${\cal D}^2_0\cong {\cal D}^2_{-1}$. Here the mapping
is the conjugation.

\subsection{Relation with cohomology}

To prove that the modules ${\cal D}^2_{-{1\over 2}}$ and
${\cal D}^2_0\cong {\cal D}^2_{-1}$ are not isomorphic to any other module,
we use the approach of the general theory of deformations (see e.g.
\cite{fuk}).

Let us denote the coefficients $-(2\lambda +1)$ and $-\lambda(\lambda +1)$ by
$\alpha _1$ and $\alpha _2$ respectively. Take $\alpha _1$ and $\alpha _2$
as {\it independent parameters}. One gets a 2-parameter family of actions of
$\Vect (M)$ more general than the action (\ref{11}) on the space
of differential operators:

\proclaim Lemma 4. For each values of $\alpha_1,\alpha_2$, the following
expression defines a $\Vect (M)$-action:
\begin{equation}
\left\{
\matrix{
T^{\alpha _1\alpha _2}_{\xi }(a_2)^{ij}
&=&
(L_{\xi }a_2)^{ij} \hfill\cr\noalign{\smallskip}
T^{\alpha _1\alpha _2}_{\xi }(a_1)^l
&=&
(L_{\xi }a_1)^l
+\alpha _1a_2^{ij}\partial_i\partial_j\xi^l\hfill\cr\noalign{\smallskip}
T^{\alpha _1\alpha _2}_{\xi }(a_0)\hfill
&=&
L_{\xi }a_0
+\alpha_2\partial_i(a_2^{jm})\partial_j\partial_m\xi^i\hfill\cr\noalign{\smallskip}
}\right.
\label{13}
\end{equation}\par

{\bf Proof}. This fact is evident since the formula (\ref{11}) defines an
action of $\Vect (M)$ and the coefficients denoted by
$\alpha_1$ and $\alpha_2$ are independent.

\vskip 0,3cm

The action (\ref{13}) is a {\it 2-parameter deformation} of the standard
$\Vect (M)$-module structure on the space $S^2(M)\oplus \Vect (M)\oplus
C^{\infty }(M)$.

\proclaim Lemma 5. The following two mappings:
$$
C_1:\Vect (M)\to\Hom(S^2(M), \Vect (M))
$$
$$
C_2:\Vect (M)\to\Hom(S^2(M), C^{\infty}(M))
$$
given by:
$
C_1(\xi)(A) = a_2^{ij}\partial_i\partial_j(\xi^l)\partial_l
$
and
$
C_2(\xi)(A) =\partial_i(a_2^{jm})\partial_j\partial_m\xi^i
$
are 1-cocycles.\par

{\bf Proof}. One should check that for any $\xi,\eta\in\Vect (M)$ :
$$
[L_{\xi},C(\eta)]-[L_{\eta},C(\xi)]=C([\xi,\eta]),
$$
whenever $C=C_1,C_2$.
This relation readily follows from the fact that
the formula (\ref{13}) defines an action of $\Vect (M)$.

\vskip 0,3cm

Standard arguments show that the structures of $\Vect (M)$-module
given by (\ref{13}) with $\alpha_1\not =0$ and with $\alpha_1=0$ are
isomorphic if and only if the cocycle $C_1$ is a coboundary. In the same way,
the  $\Vect (M)$-modules (\ref{13}) with $\alpha_2\not =0$ and with
$\alpha_2=0$ iff the cocycle $C_2$ is a coboundary.

Moreover, let us prove that it is
sufficient to study $C_1$ and $C_2$ as {\it differentiable} cocycles. This
means that we consider the groups:
$$
H^1_{\Delta }(\Vect (M);\Hom (S^2(M), \Vect (M)))
$$
and
$$
H^1_{\Delta }(\Vect (M);\Hom (S^2(M), C^{\infty}(M)))
$$
of {\it differentiable (or diagonal) cohomology}.
In other words, it is sufficient to consider the cohomology classes of $C_1$
and $C_2$ only modulo coboundaries given by differential operators (see
\cite{fuk} for details).

\proclaim Lemma 5. If $C_1$ and $C_2$ represent nontrivial classes of
differentiable cohomology, then there exists three non-isomorphic
structures of \allowbreak $\Vect(M)$-module given by (\ref{13}):

1) $\alpha_1,\alpha_2\not =0$,

2) $\alpha_1=0,\alpha_2\not =0$,

3) $\alpha_1\not =0,\alpha_2=0$.
\hfill\break

{\bf Proof}. Suppose that the $\Vect (M)$-module structures 1) and 2) are
isomorphic. Then the cocycle $C_1$ is a coboundary: there exists an operator
$B\in\Hom (S^2(M), \Vect (M))$ such that $C_1(\xi)=L_{\xi}\circ B-B\circ
L_{\xi}$. Moreover, the isomorphism  between the $\Vect (M)$-modules  1) and 2)
is given by a {\it differential operator}: $A\mapsto J(A)$. In fact any such
isomorphism is local: ${\rm supp}\,J(A)={\rm supp}\,A$. Thus, $B$ is also a
differential operator and $C_1$ is a coboundary as a differentiable cocycle.

In the same way one proves that if $C_2$ is nontrivial as a differentiable
cocycle, then the $\Vect (M)$-modules 1) and 2) are not isomorphic to the
module 3). The lemma is proven.

\subsection{Nontrivial cohomology class in \newline
$H_{\Delta}^1(\Vect (M);\Hom (S^2(M),\Vect (M)))$}

\proclaim Proposition 2. (i) If $\dim M\geq 2$, then the cocycle $C_1$
represents a  nontrivial cohomology class of the differentiable cohomology
group
$$
H_{\Delta}^1(\Vect (M);\Hom (S^2(M),\Vect (M))).
$$
(ii) If $\dim M=1$, then $C_1$ is a coboundary.\par

{\bf Proof}. To prove the second statement, remark that in the
one-dimensional case $S^2\cong {\cal F}_2$ and
consider the following operator:
$$
B\left(a(x)(dx)^{-2}\right)={1\over 2}a'(dx)^{-1}.
$$
It is easy to check, that
$C_1(\xi)=L_{\xi}\circ B-B\circ L_{\xi}=(\delta B)(\xi)$.

Let now $\dim M\geq 2$. Suppose that there exists a differential operator
$B:S^2(M)\rightarrow \Vect (M)$ such that $\delta B(\xi)=C_1(\xi)$.
Let $a^{ij}\partial_i\otimes\partial_j\in S^2(M)$, we have in general
$$
B(a)=b^{ki_1\ldots i_m}_{ij}\partial_{i_1}\ldots \partial
_{i_m}a^{ij}\partial_k
$$
where $b^{ki_1\ldots i_m}_{ij}=b^{ki_1\ldots i_m}_{ji}$ and:\goodbreak
$$
\begin{array}{rl}
(\delta B)(\xi)a
:=&  L_{\xi}\circ B a-B\circ L_{\xi}a \hfill \\ \noalign{\smallskip}
=&\Big[
\xi^r\partial_r(b^{ki_1\ldots
i_m}_{ij}\partial_{i_1}\ldots\partial_{i_m}a^{ij})
-b^{ri_1\ldots i_m}_{ij}\partial_{i_1}\ldots\partial_{i_m}a^{ij}
\partial_r\xi^k \\ \noalign{\smallskip}
&-b^{ki_1\ldots i_m}_{ij}\partial_{i_1}\ldots \partial_{i_m}
(\xi^r\partial_ra^{ij}-a^{ir}\partial_r\xi^j-a^{jr}\partial_r\xi^i)\Big]
\partial_k.
\end{array}
$$
The condition $\delta B=C_1$ implies immediately: $m=1$, i.e. $B$ is a
first order differential operator. Indeed, the highest order (in the
derivatives of $\xi$) term in $(\delta B(\xi))(a)$ is
$$
2a^{ir}b^{ki_1\ldots i_m}_{ij}\partial_{i_1}\ldots \partial_{i_m}
\partial_r\xi^j\partial_k.
$$
{}From $\delta B=C_1$ one gets that if $m>1$, then this term vanishes for
any~$a$
and $\xi$, so $b^{ki_1\ldots i_m}_{ij}\equiv 0$.

For a first order operator $B(a)=b_{ij}^{kr}\partial_ra^{ij}\partial_k$ with
$b_{ij}^{kr}=b_{ji}^{kr}$, the
condition $\delta B=C_1$ implies:
$2b^{kr}_{ij}a^{it}\partial_r\partial_t\xi^j=
a^{rt}\partial_r\partial_t\xi^k$ for any $a$ and $\xi$ (again, we consider the
highest order term in $\xi$). One readily finds that this equation has
no solution if the dimension is $n\geq 2$. Indeed, take $\xi$ such that
$\xi^j=0$ with $j\not =j_0$. Then, $b^{kr}_{ij}=0$ if $k\not =j_0$. Comparing
this property for different values of $j_0$, one finds a contradiction:
$b^{kr}_{ij}\equiv 0$.

Proposition 2 is proven.

\vskip 0,3cm

{\bf Remark}. Define $\widetilde C_1:\Vect (M)\rightarrow \Hom (S^2(M),\Vect
(M))$ by:
$$
(\widetilde C_1(\xi))(a)=a^{kr}\partial_r\partial_i\xi^i\partial_k.
$$
Then $C_1$ turns out to be a 1-cocycle cohomologous to $\widetilde C_1$:
if \allowbreak
$B\in\Hom(S^2(M),\Vect (M))$ is given by $B(a)=\partial_ia^{ik}\partial_k$,
then $\delta B=C_1+\widetilde C_1$.

\subsection{Nontrivial cohomology class in \newline
$H_{\Delta}^1(\Vect (M);\Hom(S^2(M),C^{\infty}(M)))$}

\proclaim Proposition 3.
The cocycle $C_2$ represents a  nontrivial cohomology class of the
differentiable cohomology group
$$
H_{\Delta}^1(\Vect (M);\Hom(S^2(M),C^{\infty}(M)))
$$

{\bf Proof}. Suppose there exists a differential operator
$B:S^2(M)\rightarrow C^{\infty}(M)$ such that $\delta B(\xi)=C_2(\xi)$.
One can show that it is given by:
$B(a)=b^{i_1\ldots i_m}_{ij}\partial_{i_1}\ldots\partial_{i_m}a^{ij}.$
Then,
$$
\matrix{
(\delta B)(\xi)a & =\xi^r\partial_r
(b^{i_1\ldots i_m}_{ij}\partial_{i_1}\ldots \partial_{i_m}a^{ij})-
\hfill \cr
&-b^{i_1\ldots i_m}_{ij}\partial_{i_1}\ldots \partial_{i_m}
(\xi^r\partial_ra^{ij}-a^{ir}\partial_r\xi^j-a^{jr}\partial_r\xi^i)
\hfill \cr }
$$
where $b^{i_1\ldots i_m}_{ij}=b^{i_1\ldots i_m}_{ji}$.
The highest order term (in the derivatives of $\xi$)
in this expression is:
$2a^{ir}b^{i_1\ldots i_m}_{ij}\partial_{i_1}\ldots
\partial_{i_m}\partial_r\xi^j$.
The condition $\delta B=C_2$ implies that this term equals zero (for
any value of $m$ and for any $a$ and $\xi$)
which entails $b^{i_1\ldots i_m}_{ij}\equiv 0$.
This contradiction proves Proposition 3.

Theorem 1 is proven.

\vskip 0,3cm

{\bf Remarks}. 1. The cocycle $C_2$ is related to the coadjoint action of the
Virasoro algebra. Its group analogue is given by the Schwarzian derivative (see
\cite{kir1}).

2. Recall that, in the one-dimensional case, the cohomology group
$H^1(\Vect(M);{\cal F}_{\lambda })$ is non-trivial for $\lambda =0,-1,-2$. In
these three cases it has dimension one and is generated by the following
1-cocycles:
$$
\begin{array}{rcl}
c_0(\xi (x)\partial_x)
&=&
\xi'(x),\\ \noalign{\smallskip}
c_1(\xi (x)\partial_x)
&=&
\xi''(x)dx,\\ \noalign{\smallskip}
c_2(\xi (x)\partial_x)
&=&
\xi'''(x)(dx)^2
\end{array}
$$
respectively. Three corresponding cocycles with values in the
operator space:
$C_k:\Vect (M)\rightarrow\Hom({\cal F}_{\lambda },{\cal F}_{\lambda -k})$
are given by $(C_k(\xi))(a):=c_k(\xi)\cdot a$. It is interesting to note that
the
cocycles $C_0$ and $C_2$ are nontrivial for any value of $\lambda$, but the
cocycle $C_1$ is nontrivial only for $\lambda =0$ (cf \cite{fei}).

\section{Proof of Theorem 2.}

{\bf A. $\dim M\geq 2$}.

\vskip 0,3cm

It is easy to show that the mapping (\ref{8}) is equivariant. In fact,
it becomes specially simple in
the normal coordinate system (\ref{10}). It multiplies each normal component
by a constant:
$$
{\cal L}^2_{\mu \lambda }(\bar A)_2 = \bar a_2,
\qquad
{\cal L}^2_{\mu \lambda }(\bar A)_1 = \frac{2\mu +1}{2\lambda +1}\bar a_1,
\qquad
{\cal L}^2_{\mu \lambda }(\bar A)_0
=\frac{\mu(\mu +1)}{\lambda (\lambda+1)}\bar a_0.
$$

Let us prove the uniqueness.



\subsection{Automorphisms of the modules ${\cal D}^2_{\lambda }$}

We show in this section a remarkable property of the `critical' modules
${\cal D}^2_0$, ${\cal D}^2_{-1}$ and ${\cal D}^2_{-{1\over2}}$,
namely the existence of nontrivial automorphisms of these modules.

\goodbreak

\proclaim Proposition 4. Let $\dim M\geq 2$.
\hfill\break
(i) The modules
${\cal D}^2_{\lambda }$ have no automorphisms other than
multiplication by a constant for $\lambda \not =0,-{1\over 2},-1$.
\hfill\break
(ii) All automorphisms of the modules ${\cal D}^2_0\cong {\cal D}^2_{-1}$
and ${\cal D}^2_{-{1\over 2}}$ are proportional to (\ref{auto0}) and
(\ref{auto}) respectively.\par

{\bf Proof}.
Let ${\cal I}\in {\rm End}({\cal D}^2_{\lambda })$ be an automorphism.
This means that ${\cal I}$ commutes with the Lie derivative:
${\cal I}\circ \ad L_{\xi }-\ad L_{\xi }\circ {\cal I}=0$. We first give the
general formula for these automorphisms.

\proclaim Lemma 6. Any automorphism of ${\cal D}^2_{\lambda }$ has the
following form:
\begin{equation}
{\cal I}(A)
=
c_1a_2^{ij}\partial_i\partial_j
+
(c_2a_1^k + c_3\partial_ja_2^{kj})\partial_k
+
c_4\partial_i\partial_ja_2^{ij}
+
c_5\partial_ia_1^i+c_6a_0
\label{Ansatz}
\end{equation}
\par

{\bf Proof of the lemma}.
a)  Consider the highest order term
${\cal I}_2:{\cal D}^2_{\lambda}\rightarrow S^2(M)$:
$$
{\cal I}_2(A)={\cal I}_2^2(a_2)+{\cal I}_2^1(a_1)+{\cal I}_2^0(a_0).
$$
Then the three differential operators ${\cal I}_2^2:S^2(M)\rightarrow
S^2(M)$ and ${\cal I}_2^0:C^{\infty }(M)\rightarrow S^2(M)$ and
${\cal I}_2^1:\Vect (M)\rightarrow S^2(M)$
are {\it unar equi\-variant differential operators}, since
$[\ad L_{\xi },{\cal I}](A)=0$ for any $\xi$ and $A$ (indeed, consider
$a_2\equiv 0$ to check the invariance of  ${\cal I}_2^1$ and $a_2\equiv 0$,
$a_1\equiv 0$ to check the invariance of ${\cal I}_2^0$).
A well known theorem (see \cite{rud} and the remark in Sec.~1.3) states that
there is no such nonconstant equivariant operator. Thus,
${\cal I}_2^0={\cal I}_2^1\equiv 0$ and ${\cal I}_2$
is an operator of multiplication by a constant. Put:
${\cal I}_2(A)=c_1\cdot a_2$.

b) Consider the first order term
${\cal I}_1(A)={\cal I}_1^2(a_2)+{\cal I}_1^1(a_1)+{\cal I}_1^0(a_0)$.
Again, one obtains that ${\cal I}_1^0:C^{\infty }(M)\rightarrow\Vect (M)$ and
${\cal I}_1^1:\Vect (M)\rightarrow\Vect (M)$ are equivariant
differential operators.
Thus, one has: ${\cal I}_1^0=0$, ${\cal I}_1^1(a_1)=\const\cdot a_1$.
Put:
${\cal I}_1^1(a_1)=c_2\cdot a_1$.
The operator ${\cal I}_1^2:S^2(M)\rightarrow\Vect (M)$ verifies the following
relation:
\begin{equation}
[{\cal I}_1^2,\ad L_{\xi}](a_2)
=
\left[-(c_1-c_2)a_2^{ij}\partial_i\partial_j\xi^l+
2\lambda (c_1-c_2)a_2^{il}\partial_i\partial_k\xi^k\right]\partial_l
\label{one-two}
\end{equation}
Since the right-hand side of this equation contains only second order
derivatives of $\xi^j$, one necessarily obtains that ${\cal I}_1^2$ is a first
order differential operator, namely
${\cal I}_1^2(a_2)=\alpha^{sl}_{ij}\partial_sa_2^{ij}\partial_l$.
It follows easily from (\ref{one-two}) that
$\alpha^{sl}_{ij}=\const\cdot\delta _j^s\delta _i^l$. Finally,
${\cal I}_1^2(a_2)^l=c_3\partial_ja_2^{lj}$.

c) In the same way, for the last term
${\cal I}_0(A)={\cal I}_0^2(a_2)+{\cal I}_0^1(a_1)+{\cal I}_0^0(a_0)$,
one gets: ${\cal I}_0^0(a_0)=\const\cdot a_0$,
${\cal I}_0^1(a_1)=\const\cdot\partial_ka_1^k$
and
${\cal I}_0^2(a_2)=\const\cdot\partial_i\partial_ja_2^{ij}$.
Lemma 6 is proven.

\medskip


To finish the proof of Proposition 4, substitute the expression
(\ref{Ansatz}) for ${\cal I}$ into the equation $[\ad L_{\xi },{\cal I}]$.
The result reads:
$$
\left\{
\begin{array}{rcl}
[\ad L_{\xi},{\cal I}](A)_2^{ij}
&=&
0 \\ \noalign{\smallskip}
[\ad L_{\xi},{\cal I}](A)_1^l
&=&
(c_2+c_3-c_1)a_2^{ij}\partial_i\partial_j\xi^l \\ \noalign{\smallskip}
&&+(c_3+2\lambda (c_1-c_2))a_2^{li}
\partial_i\partial_k\xi^k \\ \noalign{\smallskip}
[\ad L_{\xi},{\cal I}](A)_0
&=&
(c_5-\lambda (c_2- c_6))a_1^r\partial_r\partial_k\xi^k \\ \noalign{\smallskip}
&&
+(2c_4+(1+2\lambda)c_5
+\lambda (c_6+c_1))a_2^{ij}
\partial_i\partial_j\partial_k\xi^k \\ \noalign{\smallskip}
&&
+(2c_4+2\lambda c_5-\lambda c_3)
\partial_i(a_2^{ir})\partial_r\partial_k\xi^k \\ \noalign{\smallskip}
&&
+(c_4+c_5)\partial_r(a_2^{ij})\partial_i\partial_j\xi^r
\end{array}
\right.
$$

This expression must vanish for any $\xi$
and $A$, yielding the following conditions for the constants
$c_1,\ldots,c_6$:
$$
\left\{
\matrix{
c_2+c_3-c_1=0  \hfill \cr \noalign{\smallskip}
c_3-2\lambda (c_2-c_1)=0 \hfill \cr \noalign{\smallskip}
c_5+\lambda (c_2- c_6)=0 \hfill \cr \noalign{\smallskip}
2c_4+(1-2\lambda)c_5-\lambda (c_6-c_1)=0 \cr \noalign{\smallskip}
2c_4-2\lambda c_5+\lambda c_3=0 \hfill \cr \noalign{\smallskip}
c_4+c_5=0 \hfill \cr \noalign{\smallskip}
}
\right.
$$

If $\lambda \not =0,-{1\over 2},-1$, then this system has the
following solution: $c_1=c_2=c_6$, $c_3=c_4=c_5=0$. Thus, in this case the
modules ${\cal D}^2_{\lambda}$ have no nontrivial automorphisms.

If $\lambda =0$, then the solution is given by:
$c_1=c_2$, $c_3=c_4=c_5=0$ and
$c_6$ is a free parameter. One obtains the formula (\ref{auto0}).

If $\lambda =-{1\over 2}$, then the solution is:
$c_1=c_6$, $c_3=2c_4$, $c_2+c_3=c_6$ and $c_4+c_5=0$. This corresponds to the
formula (\ref{auto}) for the automorphisms of ${\cal D}_{-{1\over 2}}$.

Proposition 4 and Theorem 2 are proven.

\goodbreak


{\bf B. $\dim M=1$}.

\vskip 0,3cm

In the one-dimensional case, for each $\lambda \not =0,-1$, there exists a
2-parameter family of automorphisms of ${\cal D}^2_{\lambda}$. For
each value $\lambda=0,-1$ there exists a 3-parameter family. These facts
follow from the normal form (\ref{11}) of the $\Vect (M)$-action.

\section{Discussion}

\subsection{A few ideas around quantization}

May be, the most interesting
corollary of Theorem 1 is the existence of two exceptional modules of second
order differential operators: ${\cal D}^2_0$ and ${\cal D}^2_{-{1\over 2}}$.
Recall that there is no nontrivial equivariant linear mapping
${\cal D}^2_0\rightarrow {\cal D}^2_{-{1\over 2}}$.

However, these modules are of a great interest, e.g. in geometric quantization.
So, to obtain such a mapping one needs an additional structure on $M$.

\vskip 0,3cm

Given a linear connection $\Gamma^k_{ij}$ on $M$, it is possible to
define a $\Vect (M)$-equivariant linear mapping:
\begin{equation}
{\cal L}_\Gamma:
{\cal D}^2_0
\rightarrow
{\cal D}^2_{-{1\over 2}}
\label{conn}
\end{equation}
Let us give here the complete list of such mappings which are polynomial
in $\Gamma^k_{ij}$ and its partial derivatives.

\proclaim Theorem.
Let
$A
=
a_2^{ij}\partial_i\partial_j+
a_1^i\partial_i+
a_0\in{\cal D}^2_0$.
All $\Vect (M)$-equivariant linear mappings
(\ref{conn}) polynomially depending on $\Gamma ^k_{ij}$
and its partial derivatives are given by:
\begin{equation}
\label{tilde}
\left\{
\matrix{
\hfill\widetilde a_2^{ij}
&=&
a_2^{ij} \hfill\cr \noalign{\smallskip}
\hfill\widetilde a_1^i
&=&
\partial_ka_2^{ik}+c_1\nabla _ka_2^{ik}
+c_2(a_1^i+a_2^{jk}\Gamma_{jk}^i) \hfill\cr \noalign{\smallskip}
\hfill\widetilde a_0
&=&
-{1\over 4}\left(2\partial_i(a_2^{ij}\Gamma _{jk}^k)
+a_2^{ij}\Gamma _{ik}^k\Gamma _{jl}^l\right) \hfill\cr \noalign{\smallskip}
&&+c_3\nabla_i\nabla_ja_2^{ij}+
c_4\nabla _i(a_1^i+a_2^{jk}\Gamma_{jk}^i)+c_5a_2^{ij}R_{ij}+c_6a_0
}
\right.
\end{equation}
where the parameters $c_1,\ldots,c_6$ are arbitrary constants and $R_{ij}$
is the Ricci tensor.\par


{\bf Example}. When the connection is given by
a Riemannian metric $g_{ij}$ on $M$, consider the Laplace operator
$\Delta =g^{ij}\big(\partial_i\partial_j-\Gamma _{ij}^k\partial_k\big)\in
{\cal D}^2_0$.  Let $\psi\in{\cal F}_{-{1\over 2}}$ be a $1\over2$-density.
One can write locally $\psi = f\cdot (\det g)^{1\over4}$. The result of the
mapping (\ref{tilde}) is:
$$
\widetilde \Delta (f\cdot (\det g)^{1\over 4})=
(\Delta f+c_5Rf)(\det g)^{1\over 4}
$$
where $R$ is the scalar curvature.

\vskip 0,3cm

The proof of the theorem and the properties of the mapping (\ref{tilde}) will
be discussed elsewhere.

\vskip 0,3cm

{\bf Remark}. This result is consistent with the various
quantizations of the geodesic flow on Riemannian manifolds:
$-(\Delta + c R)$ where the constant $c$ is actually determined by the chosen
quantization procedure. For example, the infinitesimal
Blattner-Kostant-Sternberg pairing of real polarizations leads to
$c=-{1\over6}$ \cite{sni,woo}. In the case of $n$-dimensional spheres, the
pairing of complex polarizations gives $c=-{n-1\over4n}$ \cite{raw} while
Weyl quantization and quantum reduction would lead to $c={n+1\over4n}$ (see
\cite{det} for a survey).

\vskip 0,3cm

{\bf Remark}. So far, we have only considered linear connections; it would be
interesting to find a similar construction in terms of {\it projective
connections}. For example, in the one-dimensional case there exists a natural
mapping from ${\cal D}^2_0$ to ${\cal D}^2_{\lambda }$ using a projective
connection. Fixing a Sturm-Liouville operator $\partial^2+u(x)$ is
equivalent to fixing a projective connection.
The mapping
${\cal L}_u:{\cal D}^2_0 \rightarrow {\cal D}^2_{\lambda }$ defined by
$$
\begin{array}{rl}
{\cal L}_u(a_2\partial^2 + a_1\partial + a_0)
=
&a_2\partial^2
+
(a_1-\lambda a_2')\partial + c a_0+\lambda a_1' \\ \noalign{\smallskip}
&+\frac{1}{3}\lambda(\lambda+1)\left[a_2''-4 u a_2\right]
\end{array}
$$
is equivariant for any $c=\const$.
Indeed, the potential $u$ transforms via the Lie derivative as follows:
$L_\xi u=\xi u' + 2\xi' u + \frac{1}{2}\xi'''$ (see \cite{car}).
This construction with $\lambda ={1\over 2}, 1,
{3\over 2}, 2, \ldots$ is related to the Gelfand-Dickey bracket (see
\cite{ovs}).

\bigskip

\goodbreak

\subsection{Higher order operators}

The study of the modules of differential operators leads to first cohomology
groups
$$
H^1(\Vect (M);\Hom(S^k(M),S^m(M)))
$$
where $S^k(M)$ is the space of $k$-contravariant symmetric tensors on $M$.
Calculation of these cohomology groups seems to be a very interesting open
problem. In the one-dimensional case this problem is solved by Feigin and Fuks
in \cite{fei} for the Lie algebra of formal vector fields.

We have almost no information about $\Vect (M)$-module structures on the space
of higher order linear differential operators. Let us formulate the main
problem:

{\it Is it true that two spaces of $n$-th order differential operators on
tensor-densities are naturally isomorphic for any values of degree, except
for a finite set of critical values~?}

A positive answer would mean that there exist higher order analogues of the Lie
derivative. A negative answer would mean that second order differential
operators play a special role.

The only information that we have corresponds to the case of third order
differential operators on a one-dimensional manifold.

\proclaim Proposition. If $\dim M=1$, then the $\Vect (M)$-module structures
${\cal D}^3_{\lambda }$ on the space of differential operators
$$
a_3\partial ^3+a_2\partial ^2+a_1\partial +a_0
$$
are isomorphic to each other if $\lambda \not = 0,-1,-{1\over 2},
-{1\over 2}\pm \frac{\sqrt 21}{6}$.\par
Notice that the value $\lambda =-{1\over 2}$, absent in the case of second
order differential operators on a one-dimensional manifold, is present here.

Let us finish by two simple remarks:

1) In a particular case $\lambda +\mu =1$, the modules ${\cal
D}^n_{\lambda }$ and ${\cal D}^n_{\mu }$ are isomorphic for any $n$. The
isomorphism is given by the conjugation.

2) The module ${\cal D}^n_{-{1\over 2}}$ is not isomorphic to any module
${\cal D}^n_{\lambda }$ with $\lambda \not = -{1\over 2}$.

\vskip 1cm

{\bf Acknowledgments}.
It is a pleasure to acknowledge enlightening discussions with
A.A.~Kirillov, Y.~Kosmann-Schwarzbach, E.~JMourre and C.~Roger.


\end{document}